\documentclass[aps,prc,reprint,groupedaddress]{revtex4-1}
\usepackage{graphicx} 
\usepackage{dcolumn} 
\usepackage{bm}
\usepackage{hyperref}

\begin{document}

\title{Production of neutron-rich isotopes around N=126 in multinucleon transfer reactions}
\author{Zhao-Qing Feng}
\email{Corresponding author: fengzhq@impcas.ac.cn}

\affiliation{Institute of Modern Physics, Chinese Academy of Sciences, Lanzhou 730000, People's Republic of China}

\date{\today}
\begin{abstract}
The multinucleon transfer reactions has been investigated within the dinuclear system model. The nucleon transfer is coupled to the dissipation of relative motion energy and angular momentum by solving a set of microscopically derived master equations. A barrier distribution approach is implemented in the model in order to including the contributions of different orientations and dynamical effects. The available data of fragment production via multinucleon transfer reactions could be understood with the model for the reactions of $^{136}$Xe + $^{198}$Pt/$^{208}$Pb near Coulomb barrier energies. It is found that the production of heavy neutron-rich nuclei weakly depends on the incident energy. However, the yields of proton-rich nuclei increase with the incident energy.

\begin{description}
\item[PACS number(s)]
25.70.Hi, 24.10.Pa, 24.60.Gv
\end{description}
\end{abstract}

\maketitle

\section{Introduction}

The properties of neutron-rich heavy nuclei (N$>$100) are particularly important in understanding the nucleosynthesis during the r-process, and also related to the neutron shell evolution and to the synthesis of superheavy nuclei (SHN). However, the production of neutron-rich heavy nuclei is difficult in terrestrial laboratories via the projectile fragmentation or the asymmetric fission of actinide or superheavy nuclei. The possibility approach to reach the neutron-rich region is the multi-nucleon transfer reactions proposed by Zagrebaev and Greiner with a model based on multi-dimensional Langevin equations \cite{Za07,Za08}, in which the shell effects continue to play a significant role on the production of heavy nuclide. It is well known that the Grazing model has been successfully used for describing a few nucleon transfer reactions \cite{Wi94}. Following the motivation, several models have been developed for understanding the transfer reactions in damped heavy-ion collisions, such as the GRAZING code \cite{Graz}, dinuclear system (DNS) model \cite{Fe09a,Ad10}, dynamical model based on the Langevin equations \cite{Za15} etc. On the other hand, the microscopic approaches are proposed, i.e., the time dependent Hartree-Fock (TDHF) approach \cite{Go09,Ka13} and improved quantum molecular dynamics (ImQMD) model \cite{Ti08}.

On the experimental side, the damped collisions of two actinide nuclei were investigated at Gesellschaft f\"{u}r Schwerionenforschung (GSI) for hunting the SHN \cite{Hu77,Sc78,Kr13}. Although no evidence for SHN (Z$\geq$106), a number of actinide isotopes are produced and the yields drastically decrease with the charged number. Recently, the attempts to produce the neutron-rich nuclei around N=126 have been performed in the reactions of $^{136}$Xe+$^{208}$Pb \cite{Ko12,Ba15} and $^{136}$Xe+$^{198}$Pt \cite{Wa15}. It was found that the shell closure plays an important role on the production of neutron-rich nuclei and more advantage with the multinucleon transfer (MNT) reactions in comparison to the projectile fragmentations \cite{Ku14}. More experiments are planning for creating the heavy neutron-rich nuclei via the MNT, e.g., the heavy-ion accelerator research facility in Lanzhou (HIRFL).

In this work, the dynamics of MNT reactions in collisions of two heavy nuclei is to be investigated with the dinuclear system (DNS) model based on the statistical diffusion theory, in which a molecular configuration of two touching nuclei is assumed to keep their own individuality. The article is organized as follows. In section II we give a brief description of the DNS model and the orientation effect in the multi-nucleon transfer reactions. The reaction mechanism and calculated results are discussed in section III. Summary and perspective on the production of neutron-rich isotopes around N=126 are presented in section IV.

\section{Brief description of the model}

In the DNS model, the nucleon transfer is coupled to the relative motion by solving a set of microscopically derived master equations by distinguishing protons and neutrons \cite{Fe06,Fe07}. The time evolution of the distribution probability $P(Z_{1},N_{1},E_{1},t)$ for a DNS fragment 1 with proton number $Z_{1}$ and neutron number $N_{1}$ and with excitation energy $E_{1}$ is governed by the master equations as follows,
\begin{widetext}
\begin{eqnarray}
\frac{d P(Z_{1},N_{1},E_{1},t)}{dt}=&&\sum_{Z_{1}^{\prime
}}W_{Z_{1},N_{1};Z_{1}^{\prime},N_{1}}(t)\left[
d_{Z_{1},N_{1}}P(Z_{1}^{\prime},N_{1},E_{1}^{\prime},t)-d_{Z_{1}^{\prime
},N_{1}}P(Z_{1},N_{1},E_{1},t)\right]+\sum_{N_{1}^{\prime
}}W_{Z_{1},N_{1};Z_{1},N_{1}^{\prime}}(t) \nonumber \\
&& \left[
d_{Z_{1},N_{1}}P(Z_{1},N_{1}^{\prime},E_{1}^{\prime},t)-d_{Z_{1},N_{1}^{\prime}}P(Z_{1},N_{1},E_{1},t)\right].
\end{eqnarray}
\end{widetext}
Here the $W_{Z_{1},N_{1};Z_{1}^{\prime},N_{1}}$ ($W_{Z_{1},N_{1};Z_{1},N_{1}^{\prime}}$) is the mean transition
probability from the channel $(Z_{1},N_{1},E_{1})$ to $(Z_{1}^{\prime},N_{1},E_{1}^{\prime})$ (or $(Z_{1},N_{1},E_{1})$ to $(Z_{1},N_{1}^{\prime},E_{1}^{\prime})$), and $d_{Z_{1},N_{1}}$
denotes the microscopic dimension corresponding to the macroscopic state $(Z_{1},N_{1},E_{1})$. Sequential one-nucleon transfer is considered in the model with the relation of $Z_{1}^{\prime}=Z_{1}\pm 1$ and $N_{1}^{\prime }=N_{1}\pm 1$. The local excitation energy $E_{1}$ is determined by the dissipation energy from the relative motion and the potential energy surface of the DNS. The dissipation of the relative motion and angular momentum of the DNS is described by the Fokker-Planck equation \cite{Fe07a}. The energy dissipated into the DNS is expressed as
\begin{equation}
E^{diss}(t)=E_{c.m.}-B-\frac{\langle  J(t)\rangle(\langle J(t)\rangle+1)\hbar^{2}}{2\zeta_{rel}}-\langle  E_{rad}(J,t)\rangle.
\end{equation}
Here the $E_{c.m.}$ and $B$ are the center-of-mass energy and Coulomb barrier, respectively. The radial energy is evaluated from
\begin{equation}
\langle  E_{rad}(J,t)\rangle=E_{rad}(J,0)\exp(-t/\tau_{r})
\end{equation}
The relaxation time of the radial motion $\tau_{r}=5\times10^{-22}$s and the radial energy at the initial state $E_{rad}(J,0)=E_{c.m.}-B-J_{i}(J_{i}+1)\hbar^{2}/(2\zeta_{rel})$. The dissipation of the relative angular momentum is described by
\begin{equation}
\langle  J(t)\rangle=J_{st}+(J_{i}-J_{st})\exp(-t/\tau_{J})
\end{equation}
The angular momentum at the sticking limit $J_{st}=J_{i}\zeta_{rel}/\zeta_{tot}$ and the relaxation time $\tau_{J}=15\times10^{-22}$s. The $\zeta_{rel}$ and $\zeta_{tot}$ are the relative and total moments of inertia of the DNS, respectively. The initial angular momentum is set to be $J_{i}=J$ in the following literature.

In the relaxation process of the relative motion, the DNS will be excited by the dissipation of the relative kinetic energy. The local excitation energy is determined by the excitation energy of the composite system and the potential energy surface of the DNS. The potential energy surface (PES) of the DNS is given by
\begin{eqnarray}
U(\{\alpha\})=&&B(Z_{1},N_{1})+B(Z_{2},N_{2})-\left[B(Z,N)+V^{CN}_{rot}(J)\right]   \nonumber \\
&&+V(\{\alpha\})
\end{eqnarray}
with $Z_{1}+Z_{2}=Z$ and $N_{1}+N_{2}=N$ \cite{Fe09}. Here the symbol $\{\alpha\}$ denotes the sign of the quantities $Z_{1},N_{1}, Z_{2}, N_{2}; J, R; \beta_{1}, \beta_{2}, \theta_{1}, \theta_{2}$. The $B(Z_{i},N_{i}) (i=1,2)$ and $B(Z,N)$ are the negative binding energies of the fragment $(Z_{i},N_{i})$ and the compound nucleus $(Z,N)$, respectively, which are evaluated from the liquid drop model with including the shell and the pairing corrections. The $V^{CS}_{rot}$ is the rotation energy of the compound system. The $\beta_{i}$ represent the quadrupole deformations of the two fragments at ground state. The $\theta_{i}$ denote the angles between the collision orientations and the
symmetry axes of deformed nuclei. The interaction potential between fragment $(Z_{1},N_{1})$ and $(Z_{2},N_{2})$ includes the nuclear, Coulomb and centrifugal parts. The details are shown in Ref. \cite{Fe07}. In the calculation, the distance $R$ between the centers of the two fragments is chosen to be the value at the touching configuration, in which the DNS is assumed to be formed. So the PES depends on the proton and neutron numbers of the fragments. Shown in Fig. 1 is the driving potential (the minimum position of the PES along the mass asymmetry degree of freedom) as a function of mass asymmetry in the reaction of $^{136}$Xe+$^{208}$Pb at different orientations. The driving potential is symmetric along the mass asymmetry. It is obvious that the collision orientation plays an important role on the structure of the driving potential. Moreover, the PES dominates the yield distributions of fragments formed in transfer reactions. In this work, a barrier distribution approach is implemented into the dissipation dynamics of nucleon transfer for embodying the dynamical and orientation effects of colliding partners.

\begin{figure}
\includegraphics[width=8 cm]{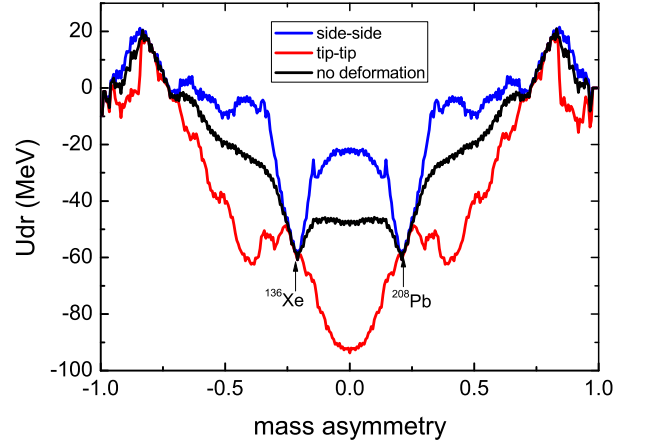}
\caption{(Color online) Driving potential as a function of mass asymmetry in the reaction of $^{136}$Xe+$^{208}$Pb at different orientations.}
\end{figure}

The cross sections of the primary fragments (Z$_{1}$, N$_{1}$) after the DNS reaches the relaxation balance are calculated as follows:
\begin{eqnarray}
\sigma_{pr}(Z_{1},N_{1},E_{c.m.})=&& \sum_{J=0}^{J_{\max}}\sigma_{cap}(E_{c.m.},J) \int  f(B)   \nonumber \\
&& \times  P(Z_{1},N_{1},E_{1},J_{1},B)dB.
\end{eqnarray}
The maximal angular momentum $J_{\max}$ is taken as the grazing collisions. The capture cross section is evaluated by $\sigma_{cap}=\pi \hbar^{2}(2J+1)/(2\mu E_{c.m.})T(E_{c.m.},J)$. The $T(E_{c.m.},J)$ is the transmission probability of the two colliding nuclei overcoming the Coulomb barrier to form a DNS and calculated with the method in Refs \cite{Fe06,Fe07}. For heavy systems, I directly use the Hill-Wheeler formula \cite{Hi53} with the barrier at the touching configuration. The survived fragments are the decay products of the primary fragments after emitting the particles and $\gamma$ rays in competition with fission. The cross sections of the survived fragments are given by
\begin{eqnarray}
\sigma_{sur}(Z_{1},N_{1},E_{c.m.})=&& \sum_{J=0}^{J_{\max}}\sigma_{cap}(E_{c.m.},J) \int  f(B)   \nonumber \\
&& \times  P(Z_{1},N_{1},E_{1},J_{1},B)   \nonumber \\
&& \times  W_{sur}(E_{1},J_{1},s)dB,
\end{eqnarray}
where the $E_{1}$ is the excitation energy of the fragment (Z$_{1}$,N$_{1}$). The maximal angular momentum is taken to be the grazing collision of two nuclei. The survival probability $W_{sur}$ of each fragment is evaluated with a statistical approach based on the Weisskopf evaporation theory \cite{Ch16}, in which the excited primary fragments are cooled by evaporating $\gamma$-rays, light particles (neutrons, protons, $\alpha$ etc) in competition with binary fission. The structure effects (shell correction, odd-even effect, Q-value etc) could be particularly significant in the formation of the primary fragments and in the decay process.

\begin{figure}
\includegraphics[width=8 cm]{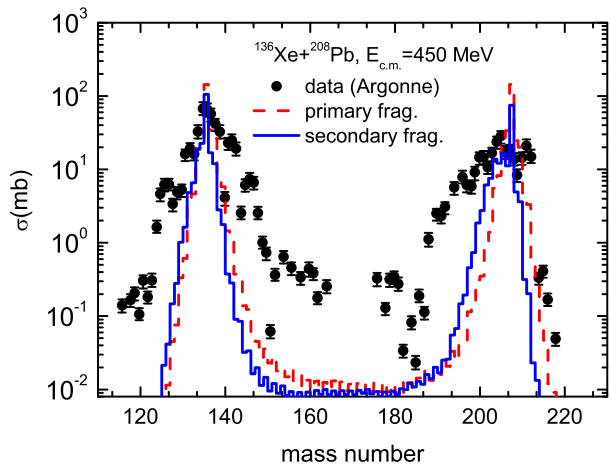}
\caption{(Color online) Mass distributions of primary and secondary fragments in the transfer reaction of $^{136}$Xe+$^{208}$Pb at the incident energy of 450 MeV and compared with the available data \cite{Ba15}.}
\end{figure}

\begin{figure}
\includegraphics[width=8 cm]{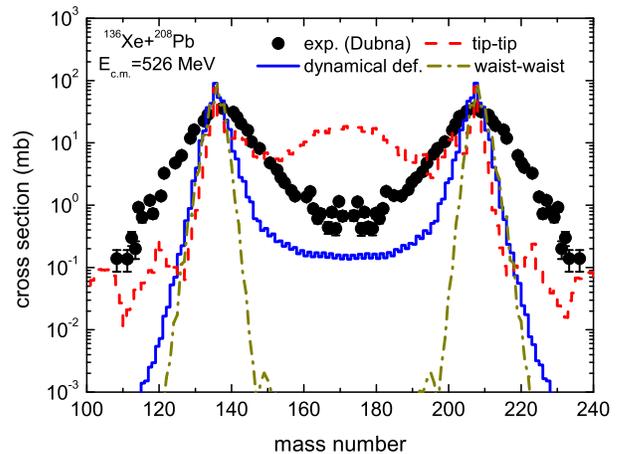}
\caption{(Color online) Orientation effects on the fragment distributions in transfer reactions and compared with the available data from Dubna \cite{Ko12}.}
\end{figure}

Shown in Fig. 2 is a comparison of the primary and survived fragments produced in the reaction $^{136}$Xe + $^{208}$Pb at the incident energy of 450 MeV. The available data were measured at the Gammasphere facility
of Argonne National Laboratory. The yields of primary fragments are symmetric along the mass asymmetry of two fragments because of the symmetric PES. Usually, production of heavy fragments (target-like) is reduced due to the comparable fission contribution. The influence of collision orientation on the fragment production is shown in Fig. 3. The structure is completely constrained by the PES. For example, the pocket shape of the driving potential in the symmetric domain in Fig. 1 leads to the increase of fragment production in the mass region of 150$\sim$190.

\section{Results and discussion}

Production cross sections of heavy or superheavy nuclei drastically decrease with the transferred nucleons, in which the isospin relaxation contributes to the fragment formation \cite{Da94}. However, the shell effect plays an important role on the production of heavy neutron-rich nuclei, i.e., isotopes around N=126 in the reaction of $^{136}$Xe+$^{208}$Pb \cite{Za08}, N=162 in collisions of $^{238}$U+$^{238}$U\cite{Fe09a}. Pure neutron transfer mechanism is investigated in the $^{136}$Xe+$^{208}$Pb reaction and compared with the data from Argonne \cite{Ba15} as shown in Fig. 4. Calculations are nicely consistent with the available experimental data at the center of mass (c.m.) energy of 450 MeV. It is interest to find that a broader distribution for the primary fragments (red lines) is formed at the energy of 526 MeV. The incident energy is close to the interaction barrier ($V_{C}=427$ MeV) at touching distance. The neutron evaporation is dominant in the decay of the primary fragments, which moves the residue nuclei towards to the neutron-poor side. Production of the heavier fragments (N$>$126) weakly depends on the incident energy because of the smaller survival probability at the higher incident energy. However, the neutron-poor nuclei (N$<$120) are related to the incident energy. Part of the secondary fragments (blue lines) in this domain are contributed from the decay of the primary fragments. Shown in Fig. 5 is the isotone production with N=126 of survived fragments in the $^{136}$Xe+$^{208}$Pb reaction at the energies of 450 MeV and 526 MeV, respectively. Production of the isotones also weakly depends on the incident energy. The distribution is not symmetric around N=126 because of the smaller neutron separation energy for the neutron-rich nuclei and more pronounced from the experimental data \cite{Ba15}. Neutron-rich nuclei of $^{204}$Pt, $^{203}$Ir and $^{202}$Os could be possibly created in experiments with cross section above 1 $nb$.

\begin{figure}
\includegraphics[width=8 cm]{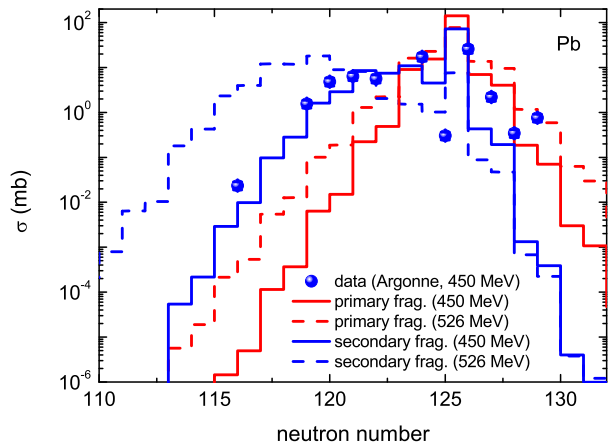}
\caption{(Color online) Production of lead isotopes in the $^{136}$Xe+$^{208}$Pb reaction at the c.m. incident energies of 450 MeV and 526 MeV, respectively.}
\end{figure}

\begin{figure}
\includegraphics[width=8 cm]{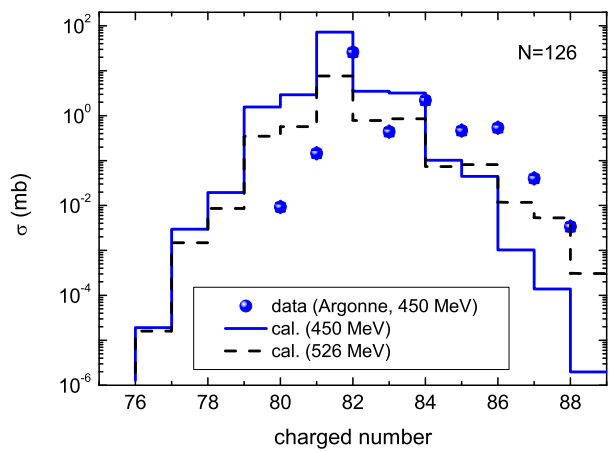}
\caption{(Color online) The same as in Fig. 4, but for the isotone distribution of N=126.}
\end{figure}

The ability of the model is further checked from the isotopic production in the $^{136}$Xe+$^{208}$Pb reaction at the c.m. energy of 450 MeV. Production of neutron-rich and neutron-poor nuclei is investigated as shown in Fig. 6 and in Fig. 7, respectively. The available data from the Argonne national laboratory \cite{Ba15} are compared with the primary (dashed lines) and survived (solid lines) fragments. Calculations show that the yields of target-like nuclei decrease with increasing the transferred protons, in particular for the Os and Rn isotopes. The odd-even and shell effects are of importance on the fragment formation. The neutron evaporation is a dominant way in cooling the excited primary fragments. The fission contribution can be negligible in the decay process because the heights of the peaks of primary and secondary fragments are similar. Besides the neutron-rich isotopes, the proton-rich nuclei can be produced via the multinucleon transfer reactions. Shown in Fig. 8 is the incident energy dependence on the production of the survived fragments in collisions of $^{136}$Xe+$^{208}$Pb. It is interest to find that the production of neutron-rich nuclei does not depend on the incident energy, e.g., Pt and Au isotopes with N$>$120, Hg and Tl with N$>$125. However, the higher energy is favorable to produce the proton-rich nuclei.

\begin{figure*}
\includegraphics[width=16 cm]{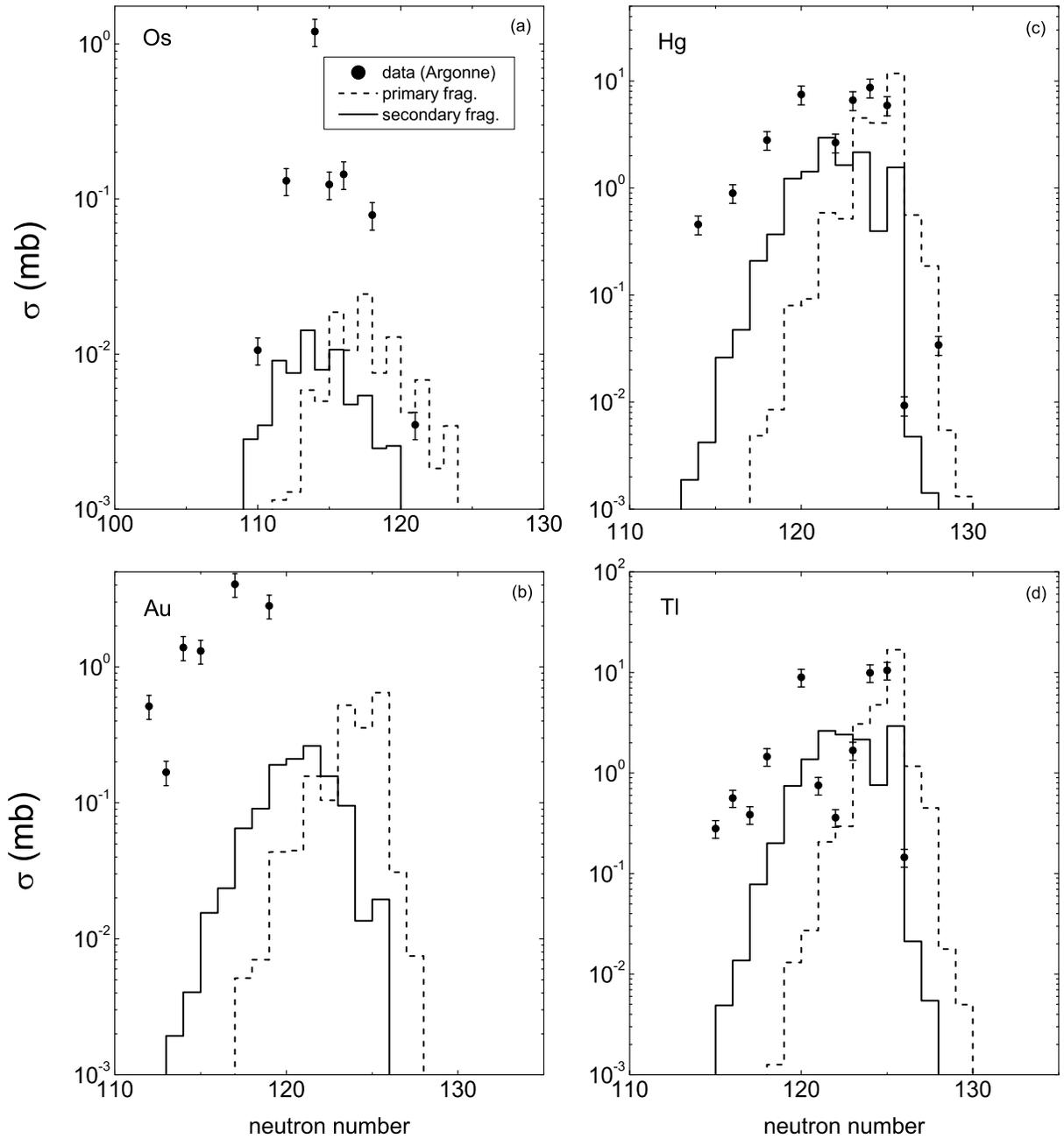}
\caption{\label{fig:wide} The isotopic production of Os, Au, Hg and Tl in the $^{136}$Xe+$^{208}$Pb reaction at the energy of 450 MeV.}
\end{figure*}

\begin{figure*}
\includegraphics[width=16 cm]{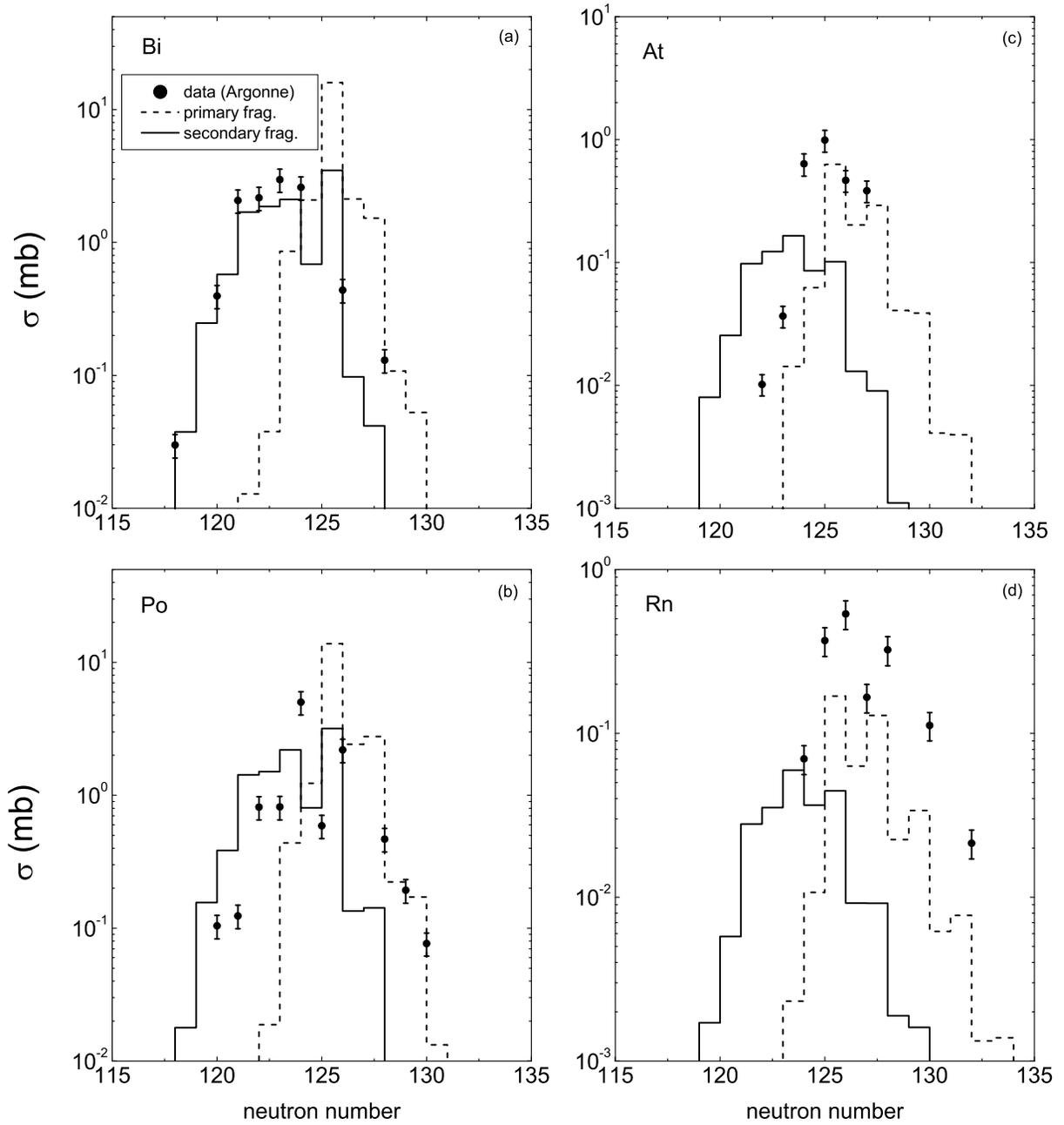}
\caption{\label{fig:wide} The same as in Fig. 6, but for the isotopes of Bi, Po, At and Rn}
\end{figure*}

\begin{figure*}
\includegraphics[width=16 cm]{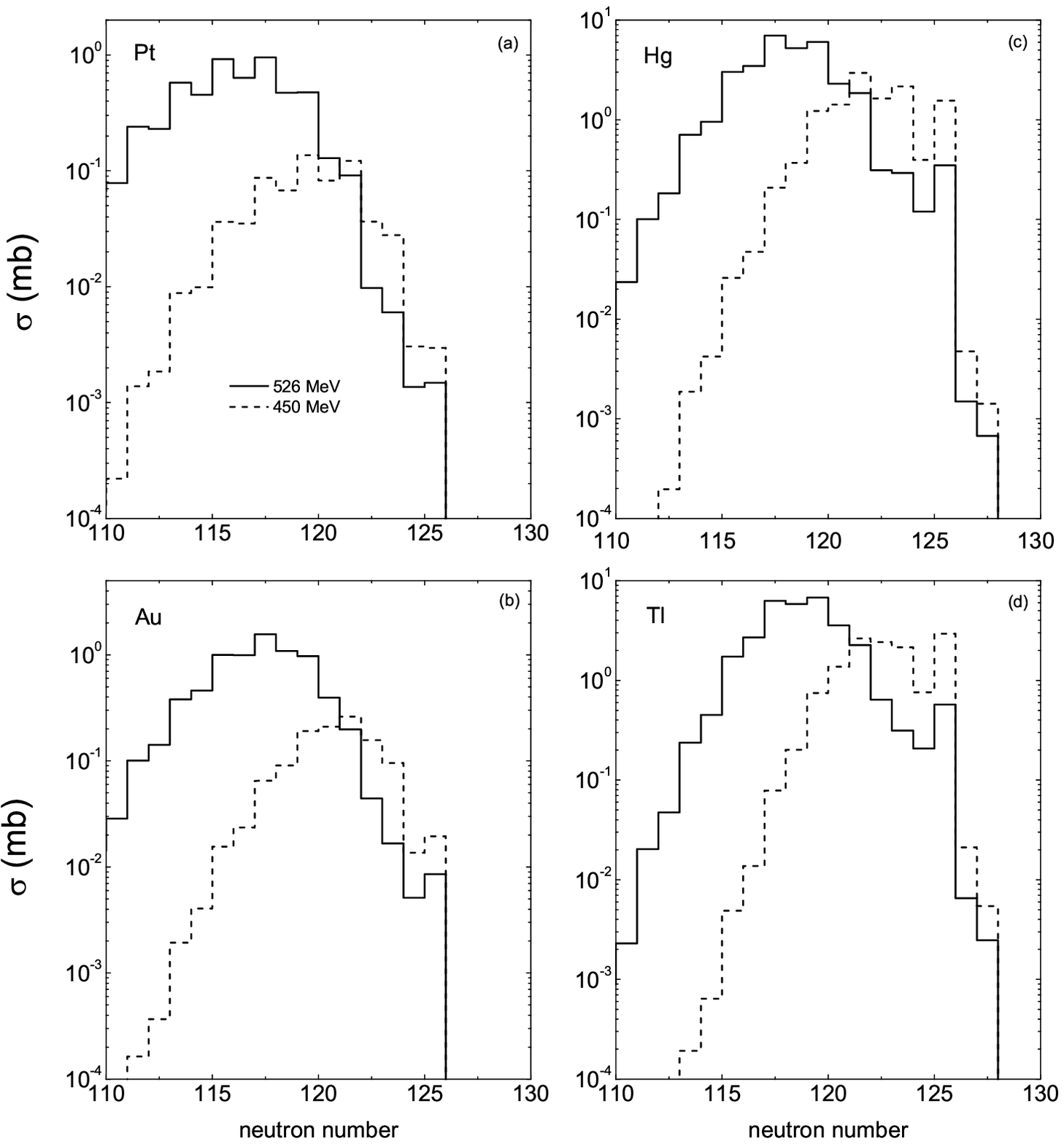}
\caption{\label{fig:wide} Incident energy dependence on the production of isotopes in the transfer reaction of $^{136}$Xe+$^{208}$Pb.}
\end{figure*}

The reaction mechanism has been further investigated with the reaction $^{136}$Xe+$^{198}$Pt. The system was attempted to produce the neutron-rich nuclei around the N=126 shell closure \cite{Wa15}. Shown in Fig. 9 is the fragment distributions formed in the MNT reactions at different bombarding energies. The production cross sections of primary fragments (red lines) are symmetric in the charge and mass spectra. It is obvious that the more nucleons are transferred with increasing the incident energy because of the more energy dissipation into the DNS. The secondary decays lead to the similar yields for heavy fragments with Z$>$85 and A$>$210 at different energies. However, the lighter projectile (target)-like fragments (A$<$130 or A$<$200) are related to the bombarding energy. It would be necessary to select the bombarding energy just above the Coulomb barrier for producing the neutron-rich heavy nuclei in the MNT reactions. The shell effect will become weak with increasing the incident energy, which is not favorable to create the neutron-rich isotopes around the neutron shell closure.

\begin{figure*}
\includegraphics[width=16 cm]{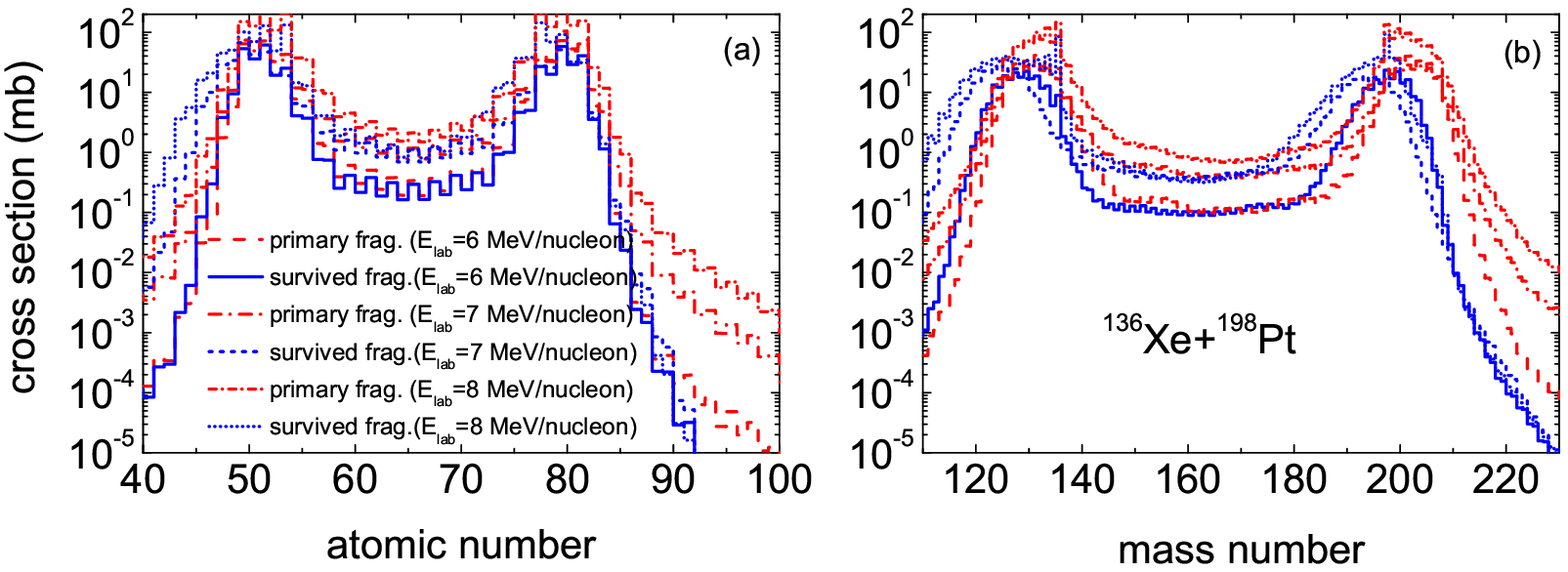}
\caption{\label{fig:wide} (Color online) Mass and charge distributions of fragments produced in multinucleon transfer reaction of $^{136}$Xe+$^{198}$Pt.}
\end{figure*}

\section{Conclusions}

In summary, the MNT mechanism in massive collisions of $^{136}$Xe on the targets of $^{198}$Pt and $^{208}$Pb near barrier energies has been investigated within the framework of the DNS model. The orientation and dynamical effects of collision systems are included in the model with the barrier distribution approach. The N=126 shell closure enables the survival of neutron-rich nuclei formed in the MNT reactions. The production of heavy nuclei weakly depends on the incident energy. The stripping nucleons increase with the bombarding energy. The isotopic yields are underestimated with increasing the transferred nucleons. Further development of the model is in progress, i.e., multi-dimensional diffusion approach with the adiabatic approximation for the PES.

\section{Acknowledgements}

This work was supported by the Major State Basic Research Development Program in China (No. 2014CB845405 and No. 2015CB856903), and the National Natural Science Foundation of China (Projects No. 11675226 and No. 11175218)

\end{document}